%% file: torsionfLM.tex
\documentclass[aps,prd,citeautoscript,twocolumn,groupedaddress,amsfonts,amssymb,amsmath,showpacs,showkeys,nofootinbib]{revtex4-1}

\usepackage{graphicx}

\usepackage{psfrag}
\usepackage{natbib}
\usepackage{setspace}

\input{aas-journals}

\begin{document}


\title{MOND as the weak field limit of an extended metric theory of gravity
with torsion}

\author{E. Barrientos}
\email[Email address: ]{ebarrientos@astro.unam.mx}
\author{S. Mendoza}
\email[Email address: ]{sergio@astro.unam.mx}
\affiliation{Instituto de Astronom\'{\i}a, Universidad Nacional
                 Aut\'onoma de M\'exico, AP 70-264, Ciudad de M\'exico, 04510,
	         M\'exico \\
            }

\date{\today}

\begin{abstract}
  In this article we construct a relativistic extended metric theory of
gravity, for which its weak field limit reduces to the non-relativistic
MOdified Newtonian Dynamics regime of gravity.  The theory is fully
covariant and local.  The way to achieve this is by introducing torsion in the
description of gravity as well as with the addition of a particular
function of the matter lagrangian into the gravitational action.
\end{abstract}


\maketitle

\section{Introduction}
\label{Introduccion}

In 1983 Milgrom proposed MOND (MOdified Newtonian Dynamics), a theory
that introduced a modification of Newton's second law in order to explain
the flattening of  rotation curves in disc galaxies \citep {Milgrom1,
Milgrom2}. From this empirical proposal it is possible to recover the
baryonic Tully-Fisher relation, for which the rotation velocity  $V$
scales as a power of the baryonic mass \( M \) (composed of stars and
dust) in the following form: $V\propto M^{1/4}$.

Although in principle, the Tully-Fisher relation was found for disc
galaxies, recent surveys have proven that it holds in dwarf spheroidal
galaxies, wide binaries and globular clusters \citep {Xavier1, Xavier2,
Xavier3, Xavier4}. Moreover, the astrophysical observations strongly
suggest that MONDian gravity accurately describes pressure supported
systems across 12 order of magnitude in mass \citep{Xavier5}.

Despite the success that MOND has at the phenomenological level,  that
formulation is non-relativistic. From a mathematical point of view,
MOND should be conceived as the weak field limit of a relativistic
proposal. Several attempts have been done towards building a relativistic
version of MOND. Amongst the many proposals in this direction we can name
the Tensor-Vector-Scalar theories \citep{Bek2, Bekenstein, Zlosnik2,
stsanders, sanders-t-v-s, SkordisTeVeS}, galileons \citep{Babichev},
bimetric theories \citep{Milgrom-bimetric}, non-local theories
\citep{Deffayet}, modified energies \citep{demir2014} and  field theories
\citep{Bruneton} to name a few.

\citet{mendozatula} built a relativistic proposal for MOND in the pure
metric formalism.  This theory is based on a dimensionally correct
action for a $f(\chi)$ function, where $\chi$ is a dimensionless
Ricci scalar defined as: $\chi=L_M^2R$, and $L_M$ is a free coupling
parameter of the theory with length dimensions, which is fixed by
recovering MOND in the weak field limit. Taking this work as starting
point \citet{barrientosmendoza} analysed the previous proposal but in
the Palatini framework.  Both,  metric and Palatini formalism yield
$\chi^{3/2}$ as the function that turns into MOND on its weak field
limit. This value is coincident with the results obtained in some
cosmological analysis \citep{SalvatoreQ1, SalvatoreQ2}.

The $f(\chi)$ theory explains not only the flattening of
rotation curves, but the correct bending angle of light for
gravitational lensing in individual, groups and clusters of galaxies
\citep{mendozalensing}. However, this proposal possess a mathematical
inconvenient since  the coupling constant $L_M$ has an explicit mass
dependence and so, it makes this proposal non-local (there
is however a mathematical way to deal with this caveat as explained
by \,\citet{carranza13,mendoza15}). As such, the $f(\chi)$ action must
not to be seen as a complete theory but as a particular case of a more
general idea.

In this article, we introduce a relativistic action, which in its weak field limit reduces
to MOND, but unlike the $f(\chi)$ theory, the coupling constants has exclusive 
dependence in pure physical constants: Newton's gravitational constant $G$,
the speed of light \( c \) and Milgrom's acceleration constant $a_0$,
making the action entirely covariant and local. This theory has two departures with respect general relativity.
On the one hand, in the geometrical sector, we work with a $f(R)$ theory with torsion. 
From the cosmological point of view, it has been proven that the torsion has 
interesting implications in order to explain the accelerated expansion of the universe 
\citep{Cosimo1, Cosimo2}. Our approach in this work is to find a MONDian
behaviour in extended metric theories of gravity with torsion.
On the other hand, based on the $f(\Sigma)$ and 
$f({\cal{L_\textrm{m}}})$ theories\citep{Harko1, Harko2, Lobo}, where $\Sigma$ is
the trace of the energy momentum tensor \( \Sigma_{\mu\nu} \) and ${\cal{L_\textrm{m}}}$ is the 
matter lagrangian, we also modify the matter sector with an action which
for this particular case is only dependent on derivatives of the matter lagrangian.

  The article is organised as follows.  Section~\ref{antecedentes}
introduces some of the theoretical background needed for torsion and for
the weak field limit of a general metric theory of gravity.  In
section~\ref{Propuestas} we present our preliminary attempts which yield
the correct MONDian proposal described in section~\ref{buena}.  Finally, in
section~\ref{discusion} we discuss our results.

\section{Background information}
\label{antecedentes}

  Before dealing with our action proposals, we first introduce some of the mathematical 
concepts which we will use throughout our work. The reader is referred to
the extensive reviews of \citet{Hehl1, Hehl2} and the summaries of
\citet{Cosimo1,Cosimo2} for further information.  As we are interested in a general 
scenario where there exists two fundamental variables, the metric $g_{\mu\nu}$ and 
a priori non-symmetric connection $\Gamma^\lambda\,_{\mu\nu}$, let us start defining 
the torsion tensor $S^\lambda\,_{\mu\nu}$  as:

\begin{equation}
	S^\lambda\,_{\mu\nu}:=\frac{1}{2}\left( \Gamma^\lambda\,_{\mu\nu}-
		\Gamma^\lambda\,_{\nu\mu}\right).
	\label{torsiondef}
\end{equation}

\noindent If we demand that this connection holds the metric compatibility
$\nabla_\lambda g_{\mu\nu}=0$, then it is possible to relate it with the Levi-Civita 
connection $\{\}^\lambda\,_{\mu\nu}$ of the metric $g_{\mu\nu}$,
through the following expression:

\begin{equation}
	\Gamma^\lambda\,_{\mu\nu}= \{\}- K^\lambda\,_{\mu\nu},
	\label{relacionGamma}
\end{equation}

\noindent where the contorsion tensor $K^\lambda\,_{\mu\nu}$  is
		given by \citep{Cosimo1}:
\begin{equation}
	K^\lambda\,_{\mu\nu}:=-S^\lambda\,_{\mu\nu}+S_{\mu\nu}\,^\lambda-S_{\nu}\,^\lambda\,_\mu.
	\label{Contorsion}
\end{equation}

The Riemann tensor is a geometric quantity defined entirely in terms of 
a general connection by: 

\begin{equation}
	R^\alpha\,_{\epsilon\mu\nu}:=\partial_\mu\Gamma^\alpha\,_{\nu\epsilon}
	-\partial_\nu\Gamma^\alpha\,_{\mu\epsilon}
	+\Gamma^\sigma\,_{\nu\epsilon} \Gamma^\alpha\,_{\mu\sigma}
	-\Gamma^\sigma\,_{\mu\epsilon} \Gamma^\alpha\,_{\nu\sigma}.
	\label{definicionriemann}
\end{equation}

\noindent Substitution of eq.\eqref{relacionGamma} into the previous 
equation yields a relation between the general  Riemann tensor 
$R^\alpha\,_{\epsilon\mu\nu}$ and the standard Riemann tensor built 
exclusively in terms of the Levi-Civita connection $R^\alpha\,_{\epsilon\mu\nu}(\{\})$:

\begin{equation}
	\begin{split}
	R^\alpha\,_{\epsilon\mu\nu}&=R^\alpha\,_{\epsilon\mu\nu}(\{\})
	+\tilde{\nabla}_\nu K^\alpha\,_{\mu\epsilon}
	-\tilde{\nabla}_\mu K^\alpha\,_{\nu\epsilon}\\
	&+K^\sigma\,_{\nu\epsilon} K^\alpha\,_{\mu\sigma}
	-K^\sigma\,_{\mu\epsilon} K^\alpha\,_{\nu\sigma}.
	\end{split}
	\label{relacionRiemann}
\end{equation}

\noindent The Ricci tensor is defined by the contraction of the first and third
index: $R_{\mu\nu}:=R^\alpha\,_{\mu\alpha\nu}$. Performing this contraction
in eq.\eqref{relacionRiemann}, yields to:

\begin{equation}
	\begin{split}
	R_{\mu\nu}&=R_{\mu\nu}(\{\})
	+\tilde{\nabla}_\nu K^\alpha\,_{\alpha\mu}
	-\tilde{\nabla}_\alpha K^\alpha\,_{\nu\mu}\\
	&+K^\sigma\,_{\nu\mu} K^\alpha\,_{\alpha\sigma}
	-K^\sigma\,_{\alpha\mu} K^\alpha\,_{\nu\sigma},
	\end{split}
	\label{relacionRi}
\end{equation}

\noindent where $ \tilde{\nabla}$ is the covariant derivative defined in 
terms of Levi-Civita connection only.

Sometimes, instead of working with the torsion tensor, it is useful to
express the results in terms of the torsion's contraction. In order to
simplify the notation we define the following tensor:

\begin{equation}
	T^\lambda\,_{\mu\nu}:=S^\lambda\,_{\mu\nu}
		+\delta^\lambda_\mu S_\nu-\delta^\lambda_\nu S_\mu,
	\label{torsionM}
\end{equation}

\noindent called the modified torsion tensor, and where 
$S_\mu:=S^\lambda\,_{\mu\lambda}$.

In this work, we use a simplified torsion term which is only vectorial 
as described in the work of \citet{torsion}.
For this particular kind of torsion, the Ricci tensor is given by: 

\begin{equation}
	R=R\{\}-2\tilde{\nabla}_\alpha T^\alpha-\frac{2}{3} T^\alpha T_\alpha.
	\label{torsionvectorial}
\end{equation}

Since we are assuming the existence of torsion, there are two 
main differences when performing the variations  and 
the use of Gauss' theorem when integrating by parts,  as compared to 
the purely metric formalism. Such differences are expressed 
in the following equations: 

\begin{gather}
	\delta R_{\mu\nu}= \nabla_\alpha \delta \Gamma^\alpha_{\nu\mu}
		-\nabla_\nu \delta \Gamma^\alpha_{\alpha\mu}
		+2 S^\alpha\,_{\lambda\nu} \delta \Gamma^\lambda_{\alpha\mu},
	\label{variacionricci}
	\\
	\intertext{and}
	\nabla_\mu {\cal{W}}^\mu=\partial_\mu {\cal{W}}^\mu
		+ 2 S^\mu\,_{\mu\nu} {\cal{W}}^\nu,
	\label{derivada}
\end{gather}
	
\noindent where $ {\cal{W}}^\mu$ is a density tensor of weight \( +1 \). 

Also, since we are interested in the non-relativistic weak field limit for 
our proposals, the metric is expanded as a Minkowskian background 
plus a small perturbation. The perturbations are given in factor terms of
order $1/c$.  For the purposes of this work, a second order
perturbation will be enough, since it this is sufficient to explain the
motion of mater and light particles at the non-relativistic level
\citep{Will}. Taking as base the work of \citep{sergioyolmo}, 
the metric coefficients at second perturbation order are given by:

\begin{equation}
  \begin{split} 
  g_{00} &= {}^{(0)}g_{00} + {}^{(2)}g_{00}=1+\frac{2\phi}{c^2},\\
  g_{ij} &= {}^{(0)}g_{ij} + {}^{(2)}g_{ij}=
  \delta_{ij}\left(-1+\frac{2\phi}{c^2}\right), \\
  g_{0i} &= 0. 
  \end{split}
\label{metricaperturbada}
\end{equation}

\section{Warming up attempts}
\label{Propuestas}

\subsection{$f(R\{\}, T)$}
\label{dos}

 Let us now make the assumption that the MONDian behavior of gravity is a physical effect due to 
the existence of torsion. The way to express this assumption is by the
addition of torsion terms to the Hilbert action. Using eq.~\eqref{torsionvectorial} as base, we propose the 
following action:

\begin{equation}
	\begin{split}
	{\cal{S}}_2&=\frac{c^3}{16\pi G L_M^2} \int \sqrt{-g} \left[
			R\{\}+\kappa\left( \tilde{\nabla}_\alpha T^\alpha+ T^\alpha T_\alpha\right)^b\right]  \, \mathrm{d}^4x\\
		&+\frac{1}{c}\int \sqrt{-g} {\cal{L_\textrm{m}}} \mathrm{d}^4x,
	  \end{split} 
\label{action2}
\end{equation}

\noindent where $\kappa$ is a coupling constant. In this case
the null variations are calculated with respect to the metric $g_{\mu\nu}$ and the 
modified torsion tensor $T^\mu$. The field equations derivated from the action~\eqref{action2}
are:

\begin{gather}
	\begin{split}
		&R_{\mu\nu}\{\}-\frac{1}{2}g_{\mu\nu}R\{\}
			-\frac{1}{2}g_{\mu\nu}\kappa(\tilde{\nabla}_\alpha
			T^\alpha+ T^\alpha T_\alpha)^b\\
		&+\kappa b(\tilde{\nabla}_\alpha T^\alpha+T^\alpha
			T_\alpha)^{b-1}T_\mu T_\nu \\ 
	        & -\kappa bT_\nu \tilde{\nabla}_\mu\left[(\tilde{\nabla}_\alpha
			T^\alpha+ T^\alpha T_\alpha)^{b-1}\right]\\
		&=\frac{8\pi G}{c^4} \Sigma_{\mu\nu},
	\end{split}
	\label{campogmunu2}
	\\
	\intertext{for the null variations with respect to the metric, and}
	2T_\mu(\tilde{\nabla}_\alpha T^\alpha+ T^\alpha T_\alpha)^{b-1}=
		\tilde{\nabla}_\mu\left[(\tilde{\nabla}_\alpha T^\alpha+ T^\alpha T_\alpha)^{b-1}\right],
	\label{campoT2}
\end{gather}

\noindent for the null variations with respect to the modified
torsion. Eq.\eqref{campoT2} is a differential equation for the torsion \(
T^\alpha \) and it can be substituted into \eqref{campogmunu2}, yielding
a single field equation:

\begin{equation}
	\begin{split}
		&R_{\mu\nu}\{\}-\frac{1}{2}g_{\mu\nu}R\{\}
			-\frac{1}{2}g_{\mu\nu}\kappa(\tilde{\nabla}_\alpha T^\alpha+ T^\alpha T_\alpha)^b\\
		&-\kappa b(\tilde{\nabla}_\alpha T^\alpha+T^\alpha T_\alpha)^{b-1}T_\mu T_\nu\\
		&=\frac{8\pi G}{c^4} \Sigma_{\mu\nu}.
	\end{split}
	\label{finalcampo2}
\end{equation}

  From the latter equation we conclude that a relation between
$R\{\}$ and $\Sigma$  is not possible because eq.\eqref{campoT2} is
a differential equation involving only $T$, and not ${\cal{L_\textrm{m}}})$.

Thus, in order to continue analysing this proposal, we need to make an extra
 assumption for the functional relation between $T$ and $\Sigma$\footnote
{By making this assumption, we are introducing additional information to
the proposal, which makes it somewhat inviable, but it will give us a good
idea on to the correct path to follow.}. Let us assume the following:

\begin{equation}
	\tilde{\nabla}_\alpha T^\alpha=0, \qquad \mbox{and} 
		\qquad T_\alpha=\kappa'\tilde{\nabla}_\alpha \Sigma,
	\label{suposiciones}
\end{equation}

\noindent where $\kappa' $ is a constant of proportionality.  At first view, it 
seems that these assumptions are very arbitrary, but the first one is for simplicity 
and the second one is based on an order of magnitude analysis 
that will recover MONDian acceleration as will be further discussed.

  With eqs.~\eqref{suposiciones}, expression~\eqref{finalcampo2} takes the
following form: 

\begin{equation}
	\begin{split}
		&R_{\mu\nu}\{\}-\frac{1}{2}g_{\mu\nu}R\{\}
			=\frac{1}{2}g_{\mu\nu}\kappa\kappa'^{2b}(\partial_ \alpha\Sigma\,\partial^\alpha\Sigma)^b\\
		&+\kappa b \kappa'^{2b}(\partial_ \alpha\Sigma \,\partial^\alpha\Sigma)^{b-1}\partial_ \mu\Sigma\,\partial_\nu\Sigma,
	\end{split}
	\label{finalcampo2.1}
\end{equation}

\noindent where we have changed $\tilde{\nabla}$ by $\partial_\alpha$ 
and dropped the Newtonian-like \( 4 \pi G \rho \) term  since we are only interested 
in the MONDian regime of gravity. Contracting the previous equation and substituting the trace of the energy-momentum 
for dust, we obtain:

\begin{equation}
	-R\{\}=(b+2)\kappa\kappa'^{2b}c^{4b}(\partial_ \alpha\rho\,\partial^\alpha\rho)^b.
	\label{trazaprop.Tpolvo}
\end{equation}

So far, we have not said anything about the constants $\kappa$ and $\kappa'$.
Due this freedom, we propose the the following constraint:

\begin{equation}
	\kappa\kappa'^{2b}c^{4b}\approx \frac{1}{c^2}.
	\label{constriccion}
\end{equation}
	
\noindent This assumption implies that to second order perturbation, the
term in parenthesis in equation~\eqref{trazaprop.Tpolvo} is a zeroth order term. For the metric \eqref{metricaperturbada}, 
the Ricci scalar at second perturbation order and the term involving the
matter density are respectively given by:

\begin{equation}
	R\{\}=-\frac{2\nabla^2\phi}{c^2} \qquad \mbox{and}
		\qquad \partial_ \alpha\rho \, \partial^\alpha\rho=-\nabla\rho\cdot\nabla\rho.
	\label{pert.prop.2}
\end{equation}

\noindent Thus, eq.\eqref{trazaprop.Tpolvo} to second perturbation order is:

\begin{equation}
	-\nabla\cdot
	\mathbf{a}=(b+2)(-1)^b\kappa\kappa'^{2b}c^{2(2b+1)}(\nabla\rho\cdot\nabla\rho)^b,
	\label{aceleracionpropuesta2}
\end{equation}

\noindent for the acceleration \( \boldsymbol{a} = -\nabla \phi \).

To order of magnitude $\rho\approx M/r^3$ and $\nabla\approx 1/r$ and so,  
the previous equation is:

\begin{equation}
	a\approx\kappa\kappa'^{2b} c^{2(2b+1)}M^{2b}r^{1-8b}.
	\label{amagnitud2}
\end{equation}

\noindent MONDian acceleration has a $ r^{-1}$ dependence. In order to
obtain that, the parameter 

\begin{equation}
	b=\frac{1}{4}.
	\label{b}
\end{equation}

\noindent With this value, the acceleration~\eqref{amagnitud2}
is given by:

\begin{equation}
	a\approx\kappa\kappa'^{1/2} c^3 M^{1/2}r^{-1}.
	\label{amagnitudb}
\end{equation}

The previous equation is important to our analysis. We have already obtained
the correct dependence on $M$ and $r$ of the MONDian acceleration. 
Therefore, the constants $\kappa$ and $\kappa'$  depend exclusively 
on $c$, $a_0$ and $G$ in the following form:

\begin{equation}
	\kappa\kappa'^{1/2}\approx \frac{(a_0 G)^{1/2}}{c^3}.
	\label{constantes}
\end{equation}

 This approach represents an entirely local and covariant relativistic 
formulation of MOND.  However, it cannot be an option to become a correct 
relativistic formulation of MOND because the
assumptions~\eqref{suposiciones} 
have no physical or mathematical support.
Despite this, the proposal gives us some clues towards the correct path to
follow in order to enhance our theory. 

\subsection{$f(R\{\}, \tilde{\nabla}_\mu  {\cal{L_\textrm{m}}}$) }
\label{tres}

 The next logical step in order to construct a relativistic formulation
of MOND consist in substituting the assumptions~\eqref{suposiciones} on 
the action \eqref{action2}. As such, we propose the following action:

\begin{equation}
	{\cal{S}}_3= \frac{16\pi G}{c^3} \int \sqrt{-g}\left[R\{\}+\lambda\tilde{\nabla}_\mu \left(  {\cal{L_\textrm{m}}}
		\tilde{\nabla}^\mu {\cal{L_\textrm{m}}}\right)\right]^\gamma \, \mathrm{d}^4x,
	\label{action3}
\end{equation}

\noindent where $\lambda$ is a coupling constant. This formulation,
unlike the two previous, has only the metric as a dynamical variable. The
field equations obtained from the null variations of the previous action
with respect to the metric are given by:

\begin{equation}
	\begin{split}
		&R_{\mu\nu}\{\}-\frac{1}{2}g_{\mu\nu} R\{\}= \frac{1}{2}g_{\mu\nu} \lambda 
			\left[\tilde{\nabla}_\alpha \left(  {\cal{L_\textrm{m}}}\tilde{\nabla}^\alpha {\cal{L_\textrm{m}}}\right)\right]^\gamma \\
		&-\gamma\lambda \left[\tilde{\nabla}_\alpha \left(  {\cal{L_\textrm{m}}}\tilde{\nabla}^\alpha {\cal{L_\textrm{m}}}\right)\right]^{\gamma-1}
			\tilde{\nabla}_\mu \left(  {\cal{L_\textrm{m}}}\tilde{\nabla}_\nu {\cal{L_\textrm{m}}}\right)\\
		&-\frac{\gamma}{2}\lambda \left( {\cal{L_\textrm{m}}} g_{\mu\nu}-\Sigma_{\mu\nu}\right) {\cal{L_\textrm{m}}}
			\tilde{\Delta}\left[\left[\tilde{\nabla}_\alpha \left(  {\cal{L_\textrm{m}}}\tilde{\nabla}^\alpha {\cal{L_\textrm{m}}}\right)\right]^{\gamma-1}\right],
	\end{split}
	\label{campoprop3}
\end{equation}

\noindent  where the Laplace-Beltrami operator $\Delta:=\nabla_\mu \nabla^\mu$. Contracting 
the latter expression with the metric $g^{\mu\nu}$ yields:

\begin{equation}
	\begin{split}
		&-R\{\}= -\lambda (\gamma-2) \left[\tilde{\nabla}_\alpha \left(  {\cal{L_\textrm{m}}}\tilde{\nabla}^\alpha {\cal{L_\textrm{m}}}\right)\right]^\gamma \\
		&-\frac{\gamma}{2}\lambda \left( 4{\cal{L_\textrm{m}}} - \Sigma \right) {\cal{L_\textrm{m}}}
			\tilde{\Delta}\left[\left[\tilde{\nabla}_\alpha \left(  {\cal{L_\textrm{m}}}\tilde{\nabla}^\alpha {\cal{L_\textrm{m}}}\right)\right]^{\gamma-1}\right].
	\end{split}
	\label{trazaprop3}
\end{equation}

  For the case of dust, the previous equation yields:

\begin{equation}
	\begin{split}
	&R\{\}=\lambda(\gamma-2)c^{4\gamma} \left[\tilde{\nabla}_\alpha \left(\rho\tilde{\nabla}^\alpha \rho\right)\right]^\gamma\\
		&+\frac{3}{2}\gamma\lambda c^{4\gamma}\rho^2 \tilde{\Delta}\left[\left[\tilde{\nabla}_\alpha \left(\rho\tilde{\nabla}^\alpha \rho \right)\right]^{\gamma-1}\right].
	\end{split}
	\label{trazapolvo3}
\end{equation}

  In order not to obtain dependence on the speed of light at second
perturbation order on the terms in between parenthesis in the previous
equation it is required that:

\begin{equation}
	\lambda c^{4\gamma}\approx \frac{1}{c^2}.
	\label{constriccion2}
\end{equation}

\noindent At the same perturbation order, 
the terms involving $\rho$ are of the zeroth
order. Using the metric \eqref{metricaperturbada}, such terms are:

\begin{equation}
	\tilde{\nabla}_\alpha \left(\rho\tilde{\nabla}^\alpha \rho\right)=-\nabla\cdot (\rho\nabla\rho)
	\qquad \mbox{and} \qquad{\tilde{\Delta} \psi}=-\nabla^2 \psi.
	\label{pert.prop.3}
\end{equation}

\noindent Direct substitution of these last two expressions and relation \eqref{pert.prop.2}, 
in eq.~\eqref{trazapolvo3} yields:

\begin{equation}
	\begin{split}
	&\frac{2\nabla^2\phi}{c^2}=-\lambda(\gamma-2)c^{4\gamma}(-1)^\gamma \left[\nabla\cdot (\rho\nabla\rho)\right]^\gamma\\
		&-\frac{3}{2}\gamma\lambda c^{4\gamma}(-1)^\gamma\rho^2 \nabla^2\left[\left[\nabla\cdot (\rho\nabla\rho)\right]^{\gamma-1}\right].
	\end{split}
	\label{tra.pol.3.orden}
\end{equation}

Based on the results of subsection~\ref{dos}, particularly on the ones in
eqs.~\eqref{b} and~\eqref{constantes}, we take the following values:

\begin{equation}
	\gamma=\frac{1}{4}, \qquad \mbox{and}
		\qquad \lambda=\zeta \frac{(a_0G)^{1/2}}{c^3},
	\label{valores}
\end{equation}

\noindent in order to obtain the following formula for the acceleration
(given by eq.\eqref{tra.pol.3.orden}):

\begin{equation}
	\begin{split}
	-\nabla\cdot\mathbf{a}&=\frac{(-1)^{1/4}}{4}\zeta (a_0G)^{1/2} \left[\left(\frac{1}{2}\nabla^2\rho^2\right)^{1/4}\right.\\
		&\left. -\frac{3}{4}\rho^2 \nabla^2 \left[\left(\frac{1}{2}\nabla^2\rho^2\right)^{-3/4}\right]\right],
	\end{split}
	\label{a3}
\end{equation}

\noindent An order of magnitude calculation of the previous equation
yields:

\begin{equation}
	a\approx \frac{(a_0 G M)^{1/2}}{r},
	\label{approx}
\end{equation}

\noindent which is the right MONDian dependence for acceleration. For
completeness, we must adjust the numerical value of $\zeta$. This is
accomplished solving analytically eq.\eqref{a3}, but this expression is
very complicated to handle and so, we will not to solve eq.\eqref{a3}
directly. Instead in the following section we put together what we have
learnt from subsections~\ref{dos} and \ref{tres}, in order to build a
theory which in its weakest field limit yields a  Poisson-like equation
less complicated than the one of~\eqref{a3}.

\section{The final proposal}
\label{buena}

\subsection{Field equations}
\label{campo}

  With all the knowledge acquired from the previous attempts, let
us start with the following action:

\begin{equation}
 	 \begin{split}
 		&{\cal{S}}_4=\omega\int \sqrt{-g} f(R)\, \mathrm{d}^4x\\
		&+\omega' \int \sqrt{-g}\left[ A\left(\tilde{\nabla}_\mu{\cal{L_\textrm{m}}}
			\tilde{\nabla}^\mu {\cal{L_\textrm{m}}}\right)^\eta+
			B\left({\cal{L_\textrm{m}}}\tilde{\Delta}{\cal{L_\textrm{m}}}\right)^\eta \right]\, \mathrm{d}^4x.
	  \end{split} 
\label{action4}
\end{equation}

\noindent where $\omega$ and $\omega'$ are the action's coupling constants.
 Since the action is a $f(R)$ function, there are two variables again, 
the connection (via torsion) and the metric. The resulting field equations are:

\begin{gather}
 	 \begin{split}
	&\omega\left( f' R_{\mu\nu} - \frac{1}{2} g_{\mu\nu}f \right)=\frac{1}{2}g_{\mu\nu}\omega'\left[A\left(\tilde{\nabla}_\mu{\cal{L_\textrm{m}}}
		\tilde{\nabla}^\mu {\cal{L_\textrm{m}}}\right)^\eta\right. \\
	&\left.+B\left({\cal{L_\textrm{m}}}\tilde{\Delta}{\cal{L_\textrm{m}}}\right)^\eta \right] 
		-A\omega' \eta\left(\tilde{\nabla}_\alpha {\cal{L_\textrm{m}}}\tilde{\nabla}^\alpha {\cal{L_\textrm{m}}}\right)^{\eta-1}
		\tilde{\nabla}_\mu{\cal{L_\textrm{m}}}\tilde{\nabla}_\nu {\cal{L_\textrm{m}}} \\
	&+A\omega' \eta \left({\cal{L_\textrm{m}}}g_{\mu\nu}-\Sigma_{\mu\nu}\right) 
		\tilde{\nabla}_\epsilon\left[\left(\tilde{\nabla}_\alpha {\cal{L_\textrm{m}}}\tilde{\nabla}^\alpha 
		{\cal{L_\textrm{m}}}\right)^{\eta-1}\tilde{\nabla}^\epsilon {\cal{L_\textrm{m}}}\right]\\
	&-B\omega'\eta\left({\cal{L_\textrm{m}}}\tilde{\Delta}{\cal{L_\textrm{m}}}\right)^{\eta-1}
		\left[{\cal{L_\textrm{m}}}\tilde{\nabla}_\mu\tilde{\nabla}_\nu {\cal{L_\textrm{m}}}
		+\frac{1}{2}\left({\cal{L_\textrm{m}}}g_{\mu\nu}-\Sigma_{\mu\nu}\right) \tilde{\Delta}{\cal{L_\textrm{m}}}\right] \\
	&-B\omega' \frac{\eta}{2}\left({\cal{L_\textrm{m}}}g_{\mu\nu}-\Sigma_{\mu\nu}\right) \tilde{\Delta}
		\left[\left({\cal{L_\textrm{m}}}\tilde{\Delta}{\cal{L_\textrm{m}}}\right)^{\eta-1}
		{\cal{L_\textrm{m}}}\right],
	\end{split}
	\label{campometrica}
	\intertext{for the null variations with respect to the metric, and:}
	\partial_\lambda f'\left( \delta^\mu_\tau \delta^\lambda_\sigma-
			\delta^\mu_\sigma \delta^\lambda_\tau\right)
			+2f' T^\mu\,_{\tau\sigma}=0,
	\label{campoconexion}
\end{gather}

\noindent for the null variations with respect to the connection and 
$f':=\partial f/\partial R$. The  corresponding  traces of the 
previous equations are given by: 

\begin{gather}
	\begin{split}
		&\omega\left( f' R-2f\right)=\omega' (2-\eta)\left[A\left(\tilde{\nabla}_\mu{\cal{L_\textrm{m}}}
			\tilde{\nabla}^\mu {\cal{L_\textrm{m}}}\right)^\eta\right.\\
		&\left.+B\left({\cal{L_\textrm{m}}}\tilde{\Delta}{\cal{L_\textrm{m}}}\right)^\eta \right] \\
		&+A\omega \eta \left(4 {\cal{L_\textrm{m}}}-\Sigma\right) \tilde{\nabla}_\epsilon
			\left[\left(\tilde{\nabla}_\alpha {\cal{L_\textrm{m}}}\tilde{\nabla}^\alpha 
			{\cal{L_\textrm{m}}}\right)^{\eta-1}\tilde{\nabla}^\epsilon {\cal{L_\textrm{m}}}\right]\\
		&-\frac{1}{2}B\omega \eta \left(4 {\cal{L_\textrm{m}}}-\Sigma\right)
			\left[\left({\cal{L_\textrm{m}}}\tilde{\Delta}{\cal{L_\textrm{m}}}\right)^{\eta-1} \tilde{\Delta}{\cal{L_\textrm{m}}}\right.\\
		&\left. \tilde{\Delta}\left[\left({\cal{L_\textrm{m}}}\tilde{\Delta}{\cal{L_\textrm{m}}}\right)^{\eta-1} {\cal{L_\textrm{m}}}\right]\right]
	\end{split}
	\label{trazametrica}
	\\
	\intertext{and:}
	\partial_\sigma f'=\frac{2}{3} f' T_\sigma.
	\label{trazaconexion}
\end{gather}

For the dust case, eq.\eqref{trazaconexion} remains the same,
while eq.\eqref{trazametrica} turns into:

\begin{equation}
	\begin{split}
		\omega\left( f' R-2f\right)&=\omega' (2-\eta)c^{4\eta}\left[A\left(\tilde{\nabla}_\mu\rho
			\tilde{\nabla}^\mu \rho\right)^\eta+B\left(\rho\tilde{\Delta}\rho\right)^\eta \right] \\
		&+3A\omega \eta c^{4\eta}\rho \tilde{\nabla}_\epsilon
			\left[\left(\tilde{\nabla}_\alpha \rho\tilde{\nabla}^\alpha 
			\rho\right)^{\eta-1}\tilde{\nabla}^\epsilon \rho\right]\\
		&-\frac{3}{2}B \omega' \eta c^{4\eta}\rho
			\left[\left(\rho\tilde{\Delta}\rho\right)^{\eta-1}\tilde{\Delta}\rho\right.\\
		&\left.+\tilde{\Delta}\left[\left(\rho\tilde{\Delta}\rho\right)^{\eta-1} \rho\right]\right]	.		
	\end{split}
	\label{trazapolvo}
\end{equation}

Let us make the following assumption:

\begin{equation}
	f(R)=R^d.
	\label{propuesta4}
\end{equation}

\noindent With this explicit relation, the traces~(eqs.\eqref{trazapolvo} 
and~\eqref{trazaconexion}) are given by:

\begin{gather}
	\begin{split}
		\omega (d-2) R^d&=\omega' (2-\eta)c^{4\eta}\left[A\left(\tilde{\nabla}_\mu\rho
			\tilde{\nabla}^\mu \rho\right)^\eta+B\left(\rho\tilde{\Delta}\rho\right)^\eta \right] \\
		&+3A\omega \eta c^{4\eta}\rho \tilde{\nabla}_\epsilon
			\left[\left(\tilde{\nabla}_\alpha \rho\tilde{\nabla}^\alpha 
			\rho\right)^{\eta-1}\tilde{\nabla}^\epsilon \rho\right]\\
		&-\frac{3}{2}B \omega' \eta c^{4\eta}\rho
			\left[\left(\rho\tilde{\Delta}\rho\right)^{\eta-1}\tilde{\Delta}\rho\right.\\
		&\left.+\tilde{\Delta}\left[\left(\rho\tilde{\Delta}\rho\right)^{\eta-1} \rho\right]\right]	
	\end{split}
	\label{trazapolvofinal}
	\\
	\intertext{and}
	T_\sigma=\frac{3}{2}(d-1)\frac{\partial_\sigma R}{R}.
	\label{conexionfinal}
\end{gather}

\subsection{MOND}
\label{valoresMOND}

Based on the results of subsections~\ref{dos} and~\ref{tres}, we choose the 
following values:

\begin{equation}
	d=4, \qquad \eta=1.
	\label{deta}
\end{equation}

\noindent Direct substitution of  these values into eqs.~\eqref{trazapolvofinal} and~\eqref{conexionfinal} 
yields:

\begin{gather}
	2\omega R^4=\omega' c^4 \left[A\tilde{\nabla}_\alpha \rho \tilde{\nabla}^\alpha \rho
		+(3A-2B)\rho\tilde{\Delta}\rho\right],
	\label{Rfinal}
	\\
	\intertext{and}
	T_\sigma=\frac{9}{2}\frac{\partial_\sigma R}{R}.
	\label{Tfinal}
\end{gather}

Let us analyse in more detail these expressions. From eq.\eqref{Rfinal} 
we obtain a relation $R=R(\rho)$ and 
substitution of this into eq.\eqref{Tfinal} yields $T=T(\rho)$.
Thus, for a vectorial torsion~\eqref{torsionvectorial} we find a relation
$R\{\}=R\{\}(\rho)$.  The end result of performing these substitutions
yields a complicated expression and so, instead 
we perform an analogous procedure to the one followed by~\citet{barrientosmendoza}
and write eq.\eqref{torsionvectorial} as:

\begin{equation}
	R=R\{\}+ H(R),
	\label{lineal1}
\end{equation}

\noindent in which we have used eq.\eqref{Tfinal} which allow us to express express
$T_\mu=T_\mu(R)$. By performing Taylor expansion for $H(R)$,
and keeping only terms up to the linear term in R, it follows that:

\begin{equation}
	 H(R)= \vartheta R + \mathcal{O}(R^2),
	\label{Taylor}
\end{equation}

\noindent where $\vartheta$ is a constant. Substitution of this result in
eq.~\eqref{torsionvectorial} gives:

\begin{equation}
	R\{\}=\vartheta' R \qquad \mbox{where} 
		\qquad \vartheta':=1-\vartheta.
	\label{lineal2}
\end{equation}

Direct substitution of this equation into eq.~\eqref{Rfinal}
yields:

\begin{equation}
	R\{\}=\vartheta'c\left[\frac{\omega'}{2\omega}\right]^{1/4}\left[A\tilde{\nabla}_\alpha \rho \tilde{\nabla}^\alpha \rho
		+(3A-2B)\rho\tilde{\Delta}\rho\right]^{1/4}.
	\label{Rmetricag}
\end{equation}

  Since we are only interested in second order terms of \( 1 / c \), we
require that the coupling constants \( \omega \) and \( \omega' \) must satisfy the
following constraint:
  
\begin{equation}
	\left[\frac{\omega'}{\omega}\right]^{1/4}\propto \frac{1}{c^3}.
	\label{con.final}
\end{equation}

\noindent From this restriction and using eqs.\eqref{pert.prop.2} and 
\eqref{pert.prop.3}, the acceleration derived from eq.\eqref{Rmetricag}
to second perturbation order is given by:

\begin{equation}
	\nabla\cdot\mathbf{a}=\vartheta'\frac{c^3}{2^{5/4}}\left[-\frac{\omega'}{\omega}\right]^{1/4}
		\left[A\nabla \rho \cdot \nabla\rho
		+(3A-2B)\rho\nabla^2\rho\right]^{1/4},
	\label{aceleracion}
\end{equation}

\noindent which, to order of magnitude yields: 

\begin{equation}
	a\approx \left[\frac{\omega'}{\omega}\right]^{1/4}
		c^3 \frac{M^{1/2}}{r}.
	\label{magnitudf}
\end{equation}

\noindent In order to recover a MONDian acceleration, the coupling 
constants $\omega$ and $\omega'$ must satisfy the following 
condition:

\begin{equation}
	\left[\frac{\omega'}{\omega}\right]^{1/4}\propto \frac{(a_0G)^{1/2}}{c^3}.
	\label{constriccionamplia}
\end{equation}

Using Buckingham's  theorem of dimensional analysis~\citep[see
e.g.][]{sedov} with $a_0$, $G$ and $c$ as the independent
variables, it follows that:

\begin{equation}
	\omega=\Lambda \frac{c^{15}}{a_0^6 G}, 
	\qquad	
	\omega'=\Lambda'\frac{c^3 G}{a_0^4},
	\label{omegas}
\end{equation}

\noindent which satisfy the requirement  \eqref{constriccionamplia}. Defining
 $\Lambda'/\Lambda:=\Xi$ and using the previous expression for the coupling
constants, eq.\eqref{aceleracion} is:

\begin{equation}
	\nabla\cdot\mathbf{a}=\vartheta'\frac{(-\Xi)^{1/4}}{2^{5/4}}(Ga_0)^{1/2}
		\left[A\nabla \rho \cdot \nabla\rho +(3A-2B)\rho\nabla^2\rho\right]^{1/4}.
	\label{aceleracionfinal}
\end{equation}

  Since we are looking for a Poisson-like equation as simply as possible, 
we choose $A=1$ and $B=3/2$, so that 
eq.\eqref{aceleracionfinal} turns into: 

\begin{equation}
	\nabla\cdot\mathbf{a}=\vartheta\frac{(-\Xi)^{1/4}}{2^{5/4}}(Ga_0)^{1/2}
		\left[\nabla \rho \cdot \nabla\rho \right]^{1/4},
	\label{aceleracion2}
\end{equation}

Solving analytically the last relation (see appendix \ref{apendice}), 
the following value of $\Xi$ is founded: 

\begin{equation}
	\Xi=-\frac{128\pi^2}{9\vartheta'^4}.
	\label{xi}
\end{equation}

\subsection{PPN consistency}
\label{PPN}

In this analysis, we expand the metric $g_{\mu\nu}$ as:

\begin{equation}
	g_{\mu\nu} = \eta_{\mu\nu} + h_{\mu\nu},
	\label{expansion}
\end{equation}

\noindent where $\eta_{\mu\nu}=\text{diag}(1, -1,-1,-1)$
is the Minkowskian metric and $h_{\mu\nu}$ is a small 
perturbation. To first order on $h_{\mu\nu}$ (second
order in $1/c^2$), the components of the Ricci tensor 
are given by:

\begin{equation}
	{}^{(2)}R_{00}\{\}=\frac{1}{2}\nabla^2 h_{00}, 
	\qquad \text{and} \qquad
	{}^{(2)}R_{ij}\{\}=\frac{1}{2}\nabla^2 h_{ij},
	\label{riccipert}
\end{equation}

\noindent for the PPN gauge \citep[see e.g][]{Will}. 

Substituting the value $\eta=1$, the functional form $f(R)=R^4$, 
the definition of $\Xi$ and eqs.~\eqref{omegas} into the full field
eqs.~\eqref{campometrica}, the following equation is 
obtained:

\begin{equation}
	\begin{split}
	&4R^3R_{\mu\nu}-\frac{1}{2}R^4=\Xi\frac{(Ga_0)^2}{c^{12}}\left[\frac{1}{2}g_{\mu\nu}
		\left(\tilde{\nabla}_\alpha {\cal{L_\textrm{m}}}\tilde{\nabla}^\alpha {\cal{L_\textrm{m}}}\right.\right.\\
	&\left.+\frac{3}{2}{\cal{L_\textrm{m}}}\tilde{\Delta}{\cal{L_\textrm{m}}}\right)
		-\tilde{\nabla}_\mu {\cal{L_\textrm{m}}}\tilde{\nabla}_\nu{\cal{L_\textrm{m}}}
		-\frac{3}{2}{\cal{L_\textrm{m}}}\tilde{\nabla}_\mu\tilde{\nabla}_\nu{\cal{L_\textrm{m}}}\\
	&\left.-\frac{1}{2}\left({\cal{L_\textrm{m}}}g_{\mu\nu}-\Sigma_{\mu\nu}\right)\tilde{\Delta}{\cal{L_\textrm{m}}}\right],
	\end{split}
	\label{campoPPN}
\end{equation}

\noindent with a trace given by: 

\begin{equation}
	\begin{split}
	2R^4=&\Xi\frac{(Ga_0)^2}{c^{12}}\left[\tilde{\nabla}_\alpha {\cal{L_\textrm{m}}}\tilde{\nabla}^\alpha {\cal{L_\textrm{m}}}
			 +\frac{3}{2}{\cal{L_\textrm{m}}}\tilde{\Delta}{\cal{L_\textrm{m}}}\right.\\
		&\left. -\frac{1}{2}\left(4{\cal{L_\textrm{m}}}-\Sigma\right)\tilde{\Delta}{\cal{L_\textrm{m}}}\right].
	\end{split}	
	\label{trazaPPN}
\end{equation}

\noindent Using this relation in eq.\eqref{campoPPN}, the field equations are: 

\begin{equation}
	\begin{split}
		4R_{\mu\nu}&=\left(\Xi\frac{(Ga_0)^2}{c^{12}}\right)^{1/4}\left[\frac{3}{4}g_{\mu\nu}
		\left(\tilde{\nabla}_\alpha {\cal{L_\textrm{m}}}\tilde{\nabla}^\alpha {\cal{L_\textrm{m}}}\right.\right.\\
	&\left.+\frac{3}{2}{\cal{L_\textrm{m}}}\tilde{\Delta}{\cal{L_\textrm{m}}}\right)
		-\tilde{\nabla}_\mu {\cal{L_\textrm{m}}}\tilde{\nabla}_\nu{\cal{L_\textrm{m}}}
		-\frac{3}{2}{\cal{L_\textrm{m}}}\tilde{\nabla}_\mu\tilde{\nabla}_\nu{\cal{L_\textrm{m}}}\\
	&\left.-\frac{1}{2}\left({\cal{L_\textrm{m}}}g_{\mu\nu}-\Sigma_{\mu\nu}\right)\tilde{\Delta}{\cal{L_\textrm{m}}}
		-\frac{1}{8}g_{\mu\nu}(4{\cal{L_\textrm{m}}}-\Sigma)\tilde{\Delta}{\cal{L_\textrm{m}}}\right]\\
	&\times \left[\tilde{\nabla}_\alpha {\cal{L_\textrm{m}}}\tilde{\nabla}^\alpha {\cal{L_\textrm{m}}}
		 +\frac{3}{2}{\cal{L_\textrm{m}}}\tilde{\Delta}{\cal{L_\textrm{m}}}
		-\frac{1}{2}\left(4{\cal{L_\textrm{m}}}-\Sigma\right)\tilde{\Delta}{\cal{L_\textrm{m}}}\right]^{-3/4} .
	\end{split}
	\label{RmunuPPN}
\end{equation}

Based on eq.\eqref{lineal1}, we can express $R_{\mu\nu}$ as:

\begin{gather}
	R_{\mu\nu}= R_{\mu\nu}\{\}+H_{\mu\nu} (R).
	\label{LinealRmunu}
	\intertext{From eq.\eqref{Taylor}, we conclude:}
	H_{\mu\nu}= \vartheta_{\mu\nu} R,
	\label{Hmunu}
\end{gather}

\noindent where $\vartheta_{\mu\nu}$ is a second rank tensor.

Using eq.\eqref{LinealRmunu} for dust,
the $00$ component of eq.\eqref{RmunuPPN} 
at second order of approximation is given by: 

\begin{equation}
	\begin{split}
	{}^{(2)} R_{00}\{\}+ {}^{(2)}H_{00}&=\frac{3(\Xi)^{1/4}}{2^{13/4}}
		\frac{(a_0 G)^{1/2}}{c^2}\left[\tilde{\nabla}_\alpha \rho \tilde{\nabla}^\alpha\rho
		+\rho\tilde{\Delta}\rho\right]\\
	&\times \left[\tilde{\nabla}_\alpha \rho \tilde{\nabla}^\alpha\rho\right]^{-3/4},
	\end{split}
	\label{temporal}
\end{equation}

\noindent where we have used the fact that the derivatives with respect 
to the coordinate $x^0$ are of order $1/c$. Comparing this latter
equation with eq.~\eqref{aceleracion2}, we find the following relation:

\begin{equation}
	\frac{1}{2}\nabla^2 h_{00}+ {}^{(2)}H_{00} =-\frac{3}{4}\frac{\nabla^2\phi}{c^2 \vartheta'}
		+G(\phi),
	\label{zerozero}
\end{equation}

\noindent where we have already substituted eqs.\eqref{riccipert},
\eqref{pert.prop.2} and \eqref{pert.prop.3}, and define $G(\phi)$ as:

\begin{equation}
	G(\phi)=\frac{3(-\Xi)^{1/4}}{2^{13/4}}
		\frac{(a_0 G)^{1/2}}{c^2}\rho\nabla^2\rho
		\left[\nabla\rho\cdot \nabla \rho\right]^{-3/4}.
	\label{G}
\end{equation}

The explicit dependence in $\phi$ is given for the solution $\rho=\rho(\phi)$
obtained by solving eq.\eqref{aceleracion2}.

In order to be in agreement with the metric~\eqref{metricaperturbada} employed in 
our exploration examples, the following relation must hold: 
$h_{00}= 2\phi/c^2$, and so:

\begin{equation}	
	{}^{(2)}H_{00}=-\frac{\nabla^2\phi}{c^2}\left(\frac{3}{4\vartheta'}+1\right) + G(\phi).
	\label{H00}
\end{equation}

Using eqs. \eqref{riccipert} and \eqref{LinealRmunu}, the spatial
components of eq.\eqref{RmunuPPN} for dust are:

\begin{equation}
	\begin{split}
		\frac{1}{2}\nabla^2 h_{ij}+{}^{(2)}H_{ij} =&\frac{\Xi^{1/4}}{2^{5/4}}\frac{(a_0 G)^{1/2}}{c^2}
		 \left[\frac{1}{4}g_{ij}\left(3\tilde{\nabla}_\alpha \rho\tilde{\nabla}^\alpha \rho\right.\right.\\
		&\left.\left.+\rho\tilde{\Delta}\rho\right)
		-\tilde{\nabla}_i\rho\tilde{\nabla}_j\rho
		-\frac{3}{2}\rho\tilde{\nabla}_i\tilde{\nabla}_j\rho\right] \\
	&\times \left[\tilde{\nabla}_\alpha\rho\tilde{\nabla}^\alpha \rho\right]^{-3/4} ,
	\end{split}
	\label{Rij1}
\end{equation}

To handle this equation in a better way, we contract it
with $\eta^{ij}$.  Defining $H_3 := \eta^{ij}H_{ij}$
and $h_3 := \eta^{ij}h_{ij}$, eq.\eqref{Rij1} turns into:

\begin{equation}
	\begin{split}
		\frac{1}{2}\nabla^2 h_3+{}^{(2)}H_3 =&\frac{\Xi^{1/4}}{2^{5/4}}\frac{(a_0 G)^{1/2}}{c^2}
		 \left[\frac{3}{4}\left(3\tilde{\nabla}_\alpha \rho\tilde{\nabla}^\alpha \rho\right.\right.\\
		&\left.\left.+\rho\tilde{\Delta}\rho\right)
		-\tilde{\nabla}_i\rho\tilde{\nabla}^i\rho
		-\frac{3}{2}\rho\tilde{\nabla}_i\tilde{\nabla}^i\rho\right] \\
	&\times\left[\tilde{\nabla}_\alpha\rho\tilde{\nabla}^\alpha
	\rho\right]^{-3/4}.
	\end{split}
	\label{Rij2}
\end{equation}

\noindent Using eqs.~\eqref{pert.prop.2}, \eqref{pert.prop.3} and~\eqref{G}
and comparing with eq.\eqref{aceleracionfinal}, the latter expression can
be expressed as:

\begin{equation}
	\frac{1}{2}\nabla^2 h_3+{}^{(2)}H_3 =-\frac{5}{4}\frac{\nabla^2\phi}{c^2\vartheta'}- G(\phi).
	\label{H2}
\end{equation}

Since we are looking for $H_{ij}$ in order to have 
$h_{ij}= \left( 2\phi / c^2 \right) \delta_{ij}$, therefore:

\begin{equation}
	{}^{(2)}H_3 =\frac{\nabla^2\phi}{c^2}\left(3-\frac{5}{4\vartheta'}\right)-G(\phi),
	\label{H3a}
\end{equation}

\noindent and because we are working in an isotropic frame, we 
conclude that: 

\begin{equation}
	{}^{(2)}H_{ij} = -\frac{\nabla^2\phi}{c^2}\left(1-\frac{5}{12\vartheta'}\right) \delta_{ij}
		+\frac{1}{3}G(\phi)\delta_{ij}.
	\label{H3b}
\end{equation}

In order to keep the contribution of $H_{\mu\nu}$ as small
as possible, we choose  the following values:

\begin{equation}
	\vartheta'=\frac{5}{12} \qquad \text {and}
		\qquad \vartheta=\frac{7}{12},
\end{equation}

\noindent which guarantee a sufficiently small value of \( {}^{(2)}H_{ij} \)
given by the second term on the right hand side of equation~\eqref{H3b}.

\section{Discussion}
\label{discusion}

  As mentioned in the introduction, many proposals of extended theories of
gravity have been constructed.  Recently, a new approach by
\citet{verlinde} yields an estimate of the excess gravity in terms of the
baryonic mass distribution and the Hubble parameter.  In a first
astrophysical test, this approach has been able to account reasonably well 
for the expected lens signal of low redshift galaxies \citep{brouwer}.
Despite this, it is not very clear from the theoretical developments of the
theory how to apply such results to an extended system such as a cluster of
galaxies.

  From the very early stages in the introduction of  torsion onto 
gravitational phenomena, it has never been thought as to which effect it can
produce. Furthermore, it has never become clear how it can affect
standard gravitational interactions.  In this work, we have shown that if
we want to understand MONDian phenomenology in the relativistic regime, we
require to extend gravity in such a way that the functional action
\(f(R,\mathcal{L}_\text{m})\) has the following form -see eq.\eqref{action4}: 

\begin{equation}
  f(R,\mathcal{L}_\text{m}) = \omega R^4  - \omega' 
    \left[ \left(\tilde{\nabla}_\mu{\cal{L_\textrm{m}}} 
    \tilde{\nabla}^\mu {\cal{L_\textrm{m}}}\right)+
    \frac{3}{2} \left({\cal{L_\textrm{m}}} 
    \tilde{\Delta}{\cal{L_\textrm{m}}}\right) \right],
\label{last}
\end{equation}

\noindent where:

\begin{equation}
  \omega = \frac{5^4 c^{15}}{2^{15} a_0^6 G} \approx \frac{ 0.02 c^{15}}{ 
    a_0^6 G}, \qquad \text{and} 
    \qquad \omega' = \frac{ 9 \pi^2 c^3 G }{ a_0^4 }.
\label{superlast}
\end{equation}

\noindent This formalism is fully covariant and local and so, unlike many of the
previous attempts built to generalise MOND to a relativistic regime it can
be tested in many astrophysical systems, such as weak and strong lensing of
individual, groups and clusters of galaxies.  It can also be applied for a 
Friedmann-Lema\^{\i}tre-Robertson-Walker universe and test the behaviour of
the large-scale universe at the present epoch.  We intend to deal with all 
these problems elsewhere.

  The departures introduced in the matter sector of the
action \eqref{last} with respect to the classical matter action $
{\cal{L_\textrm{m}}}$, brings with it some theoretical concerns since it is
not clear that such a choice would lead e.g. to geodesic trayectories, but
this is a much broader subject to discuss in the present article.
However, the motivation of choosing this particular action comes from
the field equations at the non-relativistic level, since at this level of
approximation, the field equations can be expressed as:

\begin{equation}
\left(\nabla^2\phi\right)^4\approx \left(\nabla\rho\right)^2.
\label{aproximado}
\end{equation}

\noindent In terms of the mass $M$, the radial coordinate $r$ and the
acceleration $a$, at order of magnitude, the previous equation can be
written as: 

\begin{equation}
\left(\frac{a}{r}\right)^4\approx \left(\frac{M}{r^4}\right)^2.
\label{aproximado2}
\end{equation}

This last expression yields the correct mass and radial dependence for
the MONDian acceleration. Therefore, our choice~\eqref{last} was made in
order to recover the dependence~\eqref{aproximado2}. From the above simple
calculation, this choice is not unique and others actions can be built
in order to achieve \eqref{aproximado2}. Such actions may in principle 
contain the theoretical issues that the approach introduced in this 
work presents.
 
 The fact that the matter Lagrangian appears inside the gravitational
action contradicts the precise measurements performed on Earth and on the
solar system with respect to this fact.  As it has been noted all 
throughout the article, the MONDian
behaviour of gravity occurs at mass to lenght ratios quite different from
the characteristic ones associated to the solar system.  In this respect,
the proposal constructed in this article cannot be applied to any mass to
length ratio system similar to those of the solar system.  It can only be
applied to systems where that ratio is much less than one, in which
essentially the equivalent Newtonian gravitational acceleration is \(
\lesssim a_0 \).  It is precisely on these systems where the matter
Lagrangian will appear inside the gravitational action.

The main conclusion that we can derive from this work is that in order to
recover a MONDian acceleration from a $F(R)$ theory, derivatives of the
matter Lagrangian must be present in the field equations. The proposal of
a matter Lagrangian function appearing on the gravitational action is not
new and has been studied previously~\citep{HarkoMF, sotiriou}.   The
posibility of building similar field equations from a gravitational action
that does not involve derivatives of the matter Lagrangian and satisfies
standard conservation laws is beyond the scope of this work, but will be
studied by us in future research.

\section*{Acknowledgements}
The authors acknowlodge the referee for useful suggestions that were
added to the final version of the article.  This work was supported
by DGAPA-UNAM (IN112616) and CONACyT (CB-2014-01 No.~240512) 
grants. EB and SM acknowledge economic support from CONACyT 
(517586 and 26344).

\appendix
\section{Evaluation of the constant $\Xi$}
\label{apendice}

 Since we are making the assumption that acceleration has only a radial 
component, the spherical coordinate system is the most suitable one.
As such:

\begin{gather}
	\mathbf{a}=\alpha r^\tau \hat{\boldsymbol{r}},
	\label{esfericas}
	\\
	\intertext{with divergence:}
	\nabla\cdot\mathbf{a}= \alpha (\tau+2)r^{\tau-1}.
	\label{divergencia}
	\\
	\intertext{Also, the gradient of $\rho$ in this coordinates is given by:}
	\nabla\rho=\frac{\mathrm{d} \rho}{\mathrm{d}r} \hat{\boldsymbol{r}}.
	\label{gradiente}
\end{gather}

Squaring eq.\eqref{aceleracion2} and substituting on it 
eqs.~\eqref{divergencia} and~\eqref{gradiente} yields:

\begin{equation}
	\alpha^2(\tau+2)^2 r^{2(\tau-1)}=\vartheta'^2\left(\frac{-\Xi}{2^5}\right)^{1/2}Ga_0
		\frac{d\rho}{dr},
	\label{integral1}
\end{equation}

\noindent Integrating over $r$ the latter expression gives the 
following result:

\begin{equation}
	\frac{\alpha^2(\tau+2)^2 }{2\tau-1}r^{2\tau-1}=\vartheta'^2\left(\frac{-\Xi}{2^5}\right)^{1/2}Ga_0\rho,
	\label{integral2}
\end{equation}

\noindent with $\tau\neq 1/2$. For a point mass source, the matter
density is: $\rho={M} \delta(r) / 4\pi r^2 $. 
Using this expression and integrating over $r$, we obtain:

\begin{equation}
	\frac{\alpha^2(\tau+2)^2 }{2\tau(2\tau-1)}r^{2\tau}\bigg|^{r=\infty}_{r=0}
	=\vartheta'^2\left(\frac{-\Xi}{2^5}\right)^{1/2}\frac{GMa_0}{4\pi}\frac{1}{r^2}\bigg|_{r=0},
	\label{integral3}
\end{equation}

\noindent  with the additional condition: $\tau\neq 0$. Since $\Xi$
is just a constant, it does not depend on $r$ and so, in order that
eq.\eqref{integral3} has meaning, it is necessarily that $\tau=1$.
This value was expected because we built our theory with the requirement
that $a\approx r^{-1}$. Using all this, eq.~\eqref{integral3} can be
written as:

\begin{equation}
	-\frac{\alpha^2}{6} 
	=\vartheta'^2\left(-\frac{\Xi}{2^5}\right)^{1/2}\frac{GMa_0}{4\pi}.
	\label{integral3tau}
\end{equation}

\noindent MONDiand acceleration sets the value: $\alpha=-(GMa_0)^{1/2}$ and
so:

\begin{equation}
	-\frac{1}{6} 
	=\vartheta'^2\left(-\frac{\Xi}{2^5}\right)^{1/2}\frac{1}{4\pi}.
	\label{despejar}
\end{equation}

Algebraic  manipulation of this expression yields eq.\eqref{xi}.

\bibliographystyle{aipnum4-1}
\bibliography{torsion}

\end{document}

%% file: aas-journals.tex
\newcommand{\aap}{Astronomy and Astrophysics}

\newcommand{\jcap}{Journal of Cosmology and Astroparticle Physics}

\newcommand{\mnras}{MNRAS}

\newcommand{\physscr}{Physica Scripta}